\documentclass{article}
\usepackage[utf8]{inputenc}
\usepackage{rotating}

\usepackage{amssymb}

\usepackage{rotating}

\usepackage{pdflscape}
\usepackage{fancyhdr}

\usepackage[square,numbers]{natbib}
\fancypagestyle{plain}{

\fancyhead[R]{University of Almeria}

}

\title{Software Engineering Timeline: major areas of interest and multidisciplinary trends}
\author{Isabel M. del \'Aguila, Jos\'e del Sagrado and Joaqu\'{\i}n Ca\~nadas \\
\footnotesize{Department of Informatics, University of Almer\'{\i}a,}\\  \footnotesize{ Almer\'{\i}a 04120, Spain}}

\date{December 2019}

\begin{document}

\maketitle

\begin{abstract}
Society today cannot run without software and by extension, without Software Engineering. Since this discipline emerged in 1968, practitioners have learned valuable lessons that have contributed to current practices. Some have become outdated but many are still relevant and widely used. From the personal and incomplete perspective of the authors, this paper not only reviews the major milestones and areas of interest in the Software Engineering timeline helping software engineers to appreciate the state of things, but also tries to give some insights into the trends that this complex engineering will see in the near future.


\end{abstract}

\section{Introduction}

Computer systems have progressed extraordinarily over the last half century along with one of their core components - the software. This progress has been mirrored by people's ability to embrace it; all of us use a computer on a day-to-day basis, whether directly or indirectly. Software Engineering (SE) is tasked with fostering software development; it oversees all aspects of software production, from the early stages of system specification through to system maintenance until it comes into use \cite{sommerville2010}. 

Since the late 1970s, this knowledge area has been a subtle yet fundamental part of our daily life given that software underpins countless everyday human activities. Nevertheless, none of us are aware of its presence, nor its complexity - until, of course, it fails or crashes \cite{booch2008}.

The worldwide software industry generates a huge amount of money in revenue annually and continues to expand in scope and revenue volume \cite{osterweil2008}. As with most human disciplines, SE matured out of the necessity to deal with the various challenges encountered since its inception 50 years ago.
This has created the widest variation of tools, methods and languages of any engineering field in human history. Based on these challenges, we have constructed an SE timeline that uncovers our personal multidisciplinary trend proposal that software engineers might face in the near future.

Establishing a simile with the Oedipus's answer to the riddle of the sphinx, SE has been evolving  through several ages from
childhood to senescence, passing through adulthood (see Figure 1), although senescence has not been reached yet.
The transition from childhood to adulthood   
can be dated back to the early nineties when software development established itself as a worldwide industry.
Software applications expanded to multiple domains such as telecommunications, the military, industrial processes, and entertainment, becoming in an adult discipline. The Second Age is still with us; today, we cannot know whether the advances to come, will drive SE towards a new stage, driven by multidisciplinarity and knowledge, being these the last events drawn on the proposed timeline.

We have divided the proposed ages into several eras,  as it had done in others SE history related works \cite{brennecke1996,aguila2014}, 
whose boundaries are a bit blurred. Each era is defined by a prevailing idea (see table 1) about the main challenges characterized within it. These challenges generated new SE methods and techniques to take another step in SE evolution. 
The milestones that have been selected represent either a unifying moment or a bifurcation leading to new approaches (e.g. the first one marks the birth of SE as a discipline).


\begin{sidewaysfigure}
\centering
\rotatebox{0}{
\includegraphics[scale=0.55]{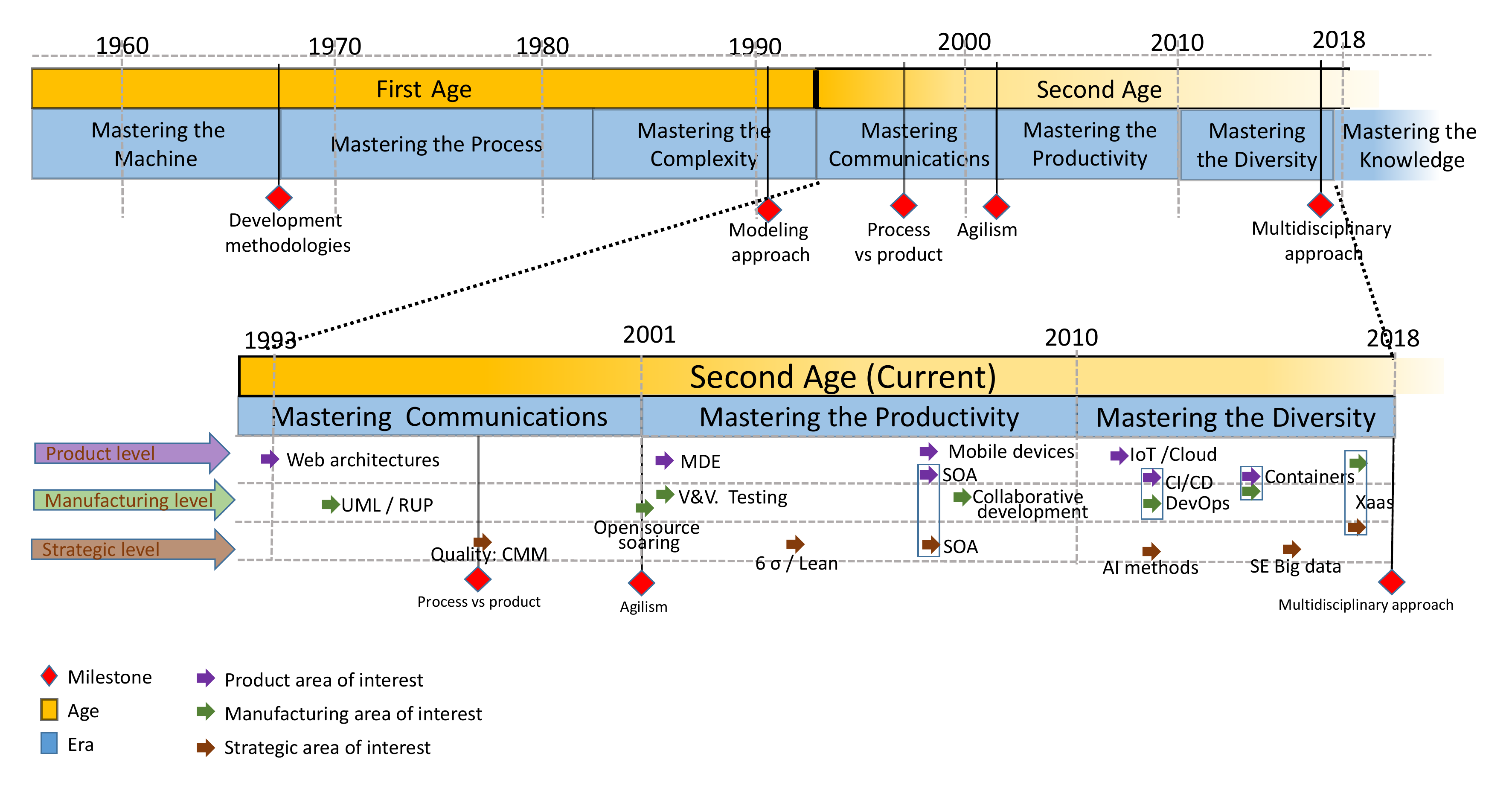}
}
\caption{SE Timeline}
\end{sidewaysfigure}


\begin{table}[ht]
\centering
\footnotesize
\caption{Prevalent idea behind each era}
\begin{tabular}{p{2ex}p{30ex}p{13ex}p{39ex}}
\hline
Age&Era & Period & This era is mastered by \\
\hline
&Mastering the Machine & 1956-1967 & Hardware resources defined software\\
1&Mastering the Process  & 1968-1982  &  Methodologies guide the software development\\
&Mastering the Complexity&1983-1992 & Domains and complexity raise force enhanced methods and tools\\ 
\hline
&Mastering Communications  & 1993-2001  & Distributed environments change the processes\\
2&Mastering the Productivity & 2001-2010 & Software factories manage the rules  \\
&Mastering the Diversity  & 2010-2017 & Devices, platforms or approaches variety expands the used methods    \\
\hline
&Mastering the Knowledge &  
2018-&SE Knowledge should be managed\\
& & ... new era?
\\
\hline
\end{tabular}
\label{eras}
\end{table}

\section{The First Age}

At the beginning, the main purpose of any software was to optimize the exploitation of the limited hardware resources available.
It was in 1956 when General Motors produced the first operating system (i.e. GM-NAA I/O for IBM704), becoming this year our starting point for our SE timeline, even when the term SE had not been coined yet. Thus,
\textbf{Mastering the Machine (1956-1967)} is the first era we have identified as the start of this Age. 
It was characterized by the lack of software development methods, which led to the origin of the term Software Engineering in NATO Science Committee, Garmisch, Germany, 1968 \cite{naur1969}
 this was the \textit{development methodologies} milestone. 

The second era, \textbf{Mastering the Process (1968-1982)}, was driven by the infamous software crisis, or maybe software chronic disease \cite{parnas1994}, 
which forced developers to focus on the stages of software specification and maintenance to deal with software aging \cite{brooks1987}.
A number of structured methods arose, such as Software Requirement Engineering Methodology (SREM)
\cite{ross1977}, or Structured Analysis and Design Technique (SADT) \cite{alford1977}, allowing the development of specification documents for business management software.
These methodologies extended the concepts of modularization and information hiding, previously applied in structured programming \cite{jackson1975,dahl1972}, from design to specification
phase.
The rise of software engineering standard is also a major accomplishment dated in this age. The community as a whole were starting to focus on standards as a means of achieving the goals \cite{TRIPP1987}

Later, in the \textbf{Mastering the Complexity era (1983-1992)}, the predominance of hardware over software came to an end, and application complexity increased exponentially. 
Computer Aided Software Engineering (CASE) tools governed this SE period, as they gave
support to engineers of this emerging discipline. Even though the main modeling approaches - data modeling and function modeling - still followed separate paths, they converged in object-oriented methods (OO) \cite{meyer1987}; 
such was the case early on with structured methodologies, which were first introduced into coding and design but eventually made their way into specification and analysis.
This OO approach enabled efficient software reuse and thus improved productivity in the process for building software \cite{meyer1987}. A second milestone, at the end of the first SE age, was the need to evolve to a \textit{modeling approach} that encouraged models to support software construction, which translated into the natural evolution of OO methods \cite{booch1986}. 
Nowadays, this is a fundamental pillar in software development. 

\section{The Second Age}

The second Age began after about 25 years after the first milestone, when SE was sufficiently resilient to reach maturity \cite{mahoney2004}. Several changes forced SE to grow up. On the one hand, software development started to be considered as an industrial business, and on the other, globalization had a deep impact on development processes, customer' feedbacks, and competitiveness.
What is more, even research community provided nurturing advances in a variety of ways to SE practitioners, although research impact was not fully felt until at least 10 years later \cite{osterweil2008}. A good example of this can be found in the work of the  software engineering coordinating committee, which began in 1997, to define  the ``generally accepted" knowledge about software engineering as a profession. This project, based on consensus, released the first trial version of the \textit{Guide to the Software Engineering Body of Knowledge (SWEBOK)} in 2001, which third  released in 2014 edition is the newest one \cite{bourque2014}.

One can delineate this second Age by describing several areas of interest that join in the timeline (see the bottom of Figure 1).
Some of them tend to set up a string of connected links evolving around similar ideas; such as the case of the thread comprising model-driven, service-oriented, containers and everything as a service areas, which exploits the notion of building software by assembling components or pieces. Figure 1 describes these areas considering three connected points of view (or levels). 
The product level includes those topics that view software as an artifact, as well as those that are physically closer to software, such as infrastructure or hardware related issues.
The manufacturing level involves the processes and methodologies employed in building software applications. 
Finally, the strategic level focuses on the business perspective, dealing with the high-level decisions and organizational tasks for the entire software development business, also called umbrella areas \cite{pressman2015}.

\subsection{Mastering Communication}
Software companies became factories in the \textbf{Mastering Communications (1993-2001)} era. The intertwining of commercial and research networks by the early 1990s marks the birth of the Internet 
as a global accessible network. On October 24, 1995, the Federal Networking Council unanimously passed a resolution defining the term Internet \cite{leiner2009brief}.
The emergence of the World Wide Web brought with it a new software concept, which causes SE methods to encompass distributed system development. \textit{Web applications and client}/\textit{server architectures} appeared as a new area of interest at the product level \cite{berners1996}. 
Object-oriented technology evolved into software reuse through the design of reusable patterns \cite{Gamma:1995:DPE:186897} and components-based software development \cite{kozaczynski1996component}.
Modeling approaches remained as a core element in software development and were included in a significant number of emerging methodologies. After this burgeoning of methods, conflicts began to appear.
Rational Corporation (now part of IBM) solved them by integrating top three methods (James Rumbaugh's, Grady Booch's and Ivar Jacobson's methods), which led to the release in 1997 of \textit{Unified Modeling Language} (UML) \cite{booch1998}.

Software factories needed to ensure quality control, which required the effective separation between process and product: \textit{the process vs product} milestone.
The \textit{Rational Unified Process} (RUP) was defined as UML partner to model the development processes \cite{jacobson99}
RUP is use case driven, architecture centric, iterative and incremental and including cycles, phases, workflows, risk mitigation, quality control, project management and configuration control. 
Certain frameworks appeared to manage both the product and the process, providing guidance for developing or improving processes to meet the business goals of an organization. These included \textit{CMM/CMMI} (capability maturity model/CMMintegrated) by the Software Engineering Institute (SEI) which adapted the
principles of process improvement from the manufacturing field to the software 
field \cite{paulk1993capability}

\subsection{Mastering the Productivity}
The fourth milestone, agilism, was a turning point in SE (agilemanifesto.org). Agile methods promoted frequent inspection and adaptation by introducing checkpoints where one can reassign customer requirements. They also encouraged software development as an incremental, cooperative, straightforward and adaptive process.
At the same time, Open-source movement propelled SE to improve the collaborative and distributed software building methods,  
Not only companies, but also developer communities started to share and enhance software applications
(www.linuxfoundation.org, www.eclipse.org). And even more, supporting knowledge also started to be shared (stackoverflow.com).
 \textit{Open source soaring} and agilism radically affected how the software had to be built, setting up the basis at manufacturing level.

Agilism ushered in a new era, \textbf{Mastering the Productivity (2001-2010)}, in which software became yet another company asset. The goals of software factories became  to reduce defects; to perform faster and more reliable processes; to increase customer satisfaction and to get greater profits. 
It also was at the start of this century, with the appearance of agilism, that one can date other two areas of interest: \textit{Model-Driven Engineering} (MDE) \cite{OMGMDA2003} 
and \textit{V\&V testing} (730-2014 - IEEE Standard for Software Quality Assurance Processes, IEEE Std 1012-2012 (Revision of IEEE Std 1012-2004) - IEEE Standard for System and Software Verification and Validation). 

MDE empowered models as first class artifacts in software development, adapting methodologies by increasing the abstraction levels of SE tasks up to those used in the problem description; enhancing productivity by the automatic execution of model transformations and code generation; and providing domain specific languages for many software development tasks. 
To address the software quality issue, verification and validation (\textit{V\&V}) methods focused on \textit{testing} technologies as a way of identifying software correctness from an agile perspective.

Although quality management methods had already been applied to SE, it is worth highlighting the use of quantitative approaches in this era, such as \textit{6$\sigma$} (Six Sigma) and \textit{Lean} \cite{biehl2004}, to support decision making at the strategic level in SE companies. 
Six Sigma, which was defined as a quality measurement to reduce variation and prevent defects, also became a management approach that promoted the need for fact-based decisions, customer focus, and teamwork. 

At the end of the decade, the widespread use of \textit{mobile devices}, together with the extended use of \textit{collaborative software development platforms} (e.g. GitHub, Jazz Project and StackOverflow),
led to \textit{service-oriented approaches} (SOA) \cite{Erl:2005:SAC:1088876} 
as a response to complexity, agile application development and software evolution that helped to improve business logic management in software industries. 
Services encapsulated data and business logic and they contain a management component, meaning that they can be perceived as units in the software assembly line.

\subsection{Mastering the Diversity}
The soaring of mobile devices, the expansion in services, and increased data availability all brought in the \textbf{Mastering the Diversity (2010-2017)} era. A wide range of widgets, from smartphones to wearables, now interconnect and exchange data continuously forming the \textit{Internet of Things (IoT)}. 
In this setting, the \textit{Cloud Computing} information technology paradigm came into play, enabling ubiquitous access to shared resources and services over the Internet. Cloud computing \cite{2010cloudcomputing} redefines how applications and services are deployed, providing scalable resources to serve customers quickly and effectively, using the required global and connected infrastructure as needed. 

\textit{Continuous Integration and Continuous Delivery (CI/CD)} \cite{cicdieeeaccess2017} provides development teams with the automation tools and techniques necessary to decrease time to market, giving a rapid software quality feedback to developers. %
The \textit{DevOps} concept,
\cite{devops2018ieee}
based on CI/CD, is considered the evolution of the agile methods thread. It focuses on the collaboration between development and operations staff throughout the development lifecycle, 
making operators to use the same techniques harnessed by developers to ensure their systems work. 
An important advance in both development and operating systems is \textit{Containers technology} 
\cite{Merkel2014contaniers}, 
which enables developers and operators to set up isolated boxes where applications are run both in development and production stages, thus reducing problems in deploying applications and services. This technology allowed software design to evolve into microservices architecture \cite{microservices}, connecting suites of independently deployable services that can work together.

The soaring levels of available services have led to the \textit{XaaS} concept (\textit{everything-as-a-service}), 
\cite{xaas2015}
which evolved as a generalization of the services provided in Cloud Computing (i.e, next element in the thread). 
XaaS became established not only as a way for providers to offer services, but also, from a strategic perspective, as a means for software companies to access up-to-date technology that is available as services, hence reducing expenditure on service consumption as well as indicating the level of cloud adoption. 
Set also the strategic level, two important areas of interest can be timelined in this era: \textit{Artificial Intelligence methods} adoption and \textit{Big Data applied to SE}. SE should deal with decision-making processes at the strategic level throughout the lifetime of a software product. Consequently, SE can be considered a knowledge-intensive process and therefore be framed within the AI domain. Furthermore, if a portion of expert knowledge was modeled and then was incorporated into the SE lifecycle (and into the tools that support it), it would greatly benefit any development process. 
Several software development and deployment processes have already seen the use of AI algorithms, such as  predictive models for software economics and risk assessment, or the search methods for finding "good-enough" solutions to large-scale SE problems caused by their computational complexity. For instance, Search Based SE \cite{sbse_HARMAN} has been applied to almost all SE activities, being software  testing the most prolific one due to their importance for collaborative development. 

A second recent ``disrupt" to SE theory and practice is the widespread availability of Big Data methods that extract valuable information from data in order to use it in intelligent ways, such as to revolutionize decision-making in businesses, science and society. This may lead to radical methods for overcoming SE problems as well as unprecedented opportunities. Huge datasets can now be stored and managed efficiently on cloud databases, providing the source for Big Data applications.
Many sources, such as forums, forges, blogs, Q\&A sites, and social networks provide a wealth of data that can be analyzed to uncover new requirements. 
They provide evidence on usage and development trends for application frameworks on which  empirical studies involving real-world software developers can be carried out \cite{miningGitHub2017}. 
In addition, real-time data collected from mobile and cloud applications is being analyzed to detect user trends, preferences and optimization opportunities.
Diversity in these emerging areas of interest has greatly impacted SE, forcing it to adopt a multidisciplinary standpoint (\textbf{the multidisciplinary approach} milestone),  where knowledge gathered from diverse disciplines should be embedded in the SE processes leading to an open SE era governed by knowledge and defining the current \textbf{Mastering the Knowledge era (2018- )}. 

\section{Multidisciplinary Trends. Towards Mastering the Knowledge era}

Given that today is yesterday's tomorrow, it is time to glimpse the movements that, in our opinion, will stay in power in the Software Engineering field during the next years. Society, software developers and leading technologies are the sources feeding the mainstream in which the three tributaries converge: connectivity, artificial intelligence and security. We strongly believe that these trends can help software engineers make advances in the software development lifecycle, either at product, manufacturing or strategic levels. Without further ado, let us discuss each of these trends in turn.

\subsection{Connectivity}
Systems are becoming more and more complex, and at an even faster rate than before. You simply need to look at the variety of objects that are now connected via wearables and appliances, from mobile devices to smart cars and homes. These devices have to be able to work properly in places with both good and bad connectivity. Not only appears the need of interconnection at gadget level, but also at software level.
Software engineers must focus on developing reliable software applications in case of temporary connection loss; development process supported by connected devices and applications cannot bring the development to a halt in case of eventual connectivity problems; and new formats and protocols for sharing data and linking behaviour between connected services will be addressed.
These are not new issues, in the same way as Model-Driven Engineering, Service-Oriented Architectures, IoT, Cloud and CI/CD are not either; indeed, connectivity has been addressed yet, at least partially, by all of them. 
However, this diversity of solutions should be approached from an SE unified point of view,  
thus, the SE challenge is to improve continuous abstraction and integration of the numerous platforms, technologies and services that will be increased in a common interconnected ecosystem. With this in mind, SE  should focus on achieving agreed faster delivery of the software quality process in response to the needs of software availability and innovation. 

\subsection{Artificial Intelligence}
Software Engineering, as any other business that wants to stay relevant, needs to adopt AI. Nowadays, data are everywhere and the software business is no exception. Data are gathered at every software stage, from requirements to maintenance. In addition, it is almost certain that for any software decision, an AI method can be found that can provide valuable help at the time of (when) making it. Search-Based Software Engineering has provided the first insights into this assertion, as it applies search-based optimization to SE lifecycle problems. Moreover, it should be pointed out that uncertainty reasoning and classification/prediction in AI have also provided assistance to software engineers in modeling software reliability and project planning, respectively. Now, in the 21st century, the big challenge is to incorporate data into software project development and to evolve towards intelligent SE automation, so that machines carry out software engineering activities as well as humans do \cite{SE2050}.  

\subsection{Security}
Recently, security flaws have been found in computer processors or their software applications that allow hackers to steal sensitive data without the users knowing, such as the soldiers' smartphones \cite{NATOtroops}.%
Other vulnerabilities have likewise affected the National Health Service and several electoral campaigns in different countries %
\cite{2018CambridgeAnalytica}. 
It is therefore essential that
, as the guarantor of software safety and quality, SE faces the challenge of ensuring that security vulnerabilities are not introduced during software development. 
The first attempts at this can be found in the Open Web Applications Security Project, where software security is assessed  and knowledge-based documentation concerning application security is issued. 
Great efforts are being carried out to automate security vulnerability checking in the software lifecycle. As example, GitHub started to report security vulnerabilities in project dependencies \cite{githubSecurity2017} 
 quite recently. However, that is only the beginning of a wider and huge effort that must be addressed in the following years. 

\section*{Acknowledgements} This research has been financed by the Spanish Ministry of Economy and Competitiveness under project TIN2016-77902-C3-3-P (PGM-SDA II project) and partially supported by Data, Knowledge and Software Engineering (DKSE) research group (TIC-181) of the University of Almer\'{\i}a, the Agrifood Campus of International Excellence (ceiA3).

\end{document}